\documentclass[aip,apl,twocolumn,superscriptaddress,reprint,amsmath,floatfix]{revtex4-1}
\usepackage{graphicx}% Include figure files
\usepackage{array}
\usepackage{dcolumn}% Aligning on a decimal point
\newcommand*{\bra}[1]{\langle #1 |}
\newcommand*{\ket}[1]{| #1 \rangle}

\begin{document}

\title{Conquer the fine structure splitting of excitons in self-assembled
InAs/GaAs quantum dots via combined stresses}

\author{Jianping Wang}
\affiliation{Key Laboratory of Quantum Information, University of Science and
Technology of China, Hefei, Anhui, 230026, People's Republic of China}

\author{Ming Gong}
\affiliation{Department of Physics and Astronomy, Washington State University,
Pullman, Washington, 99164 USA}

\author{Guang-Can Guo}
\affiliation{Key Laboratory of Quantum Information, University of Science and
Technology of China, Hefei, Anhui, 230026, People's Republic of China}

\author{Lixin He}
\email{helx@ustc.edu.cn}
\affiliation{Key Laboratory of Quantum Information, University of Science and
Technology of China, Hefei, Anhui, 230026, People's Republic of China}

\date{\today}
%\keywords{Quantum dot, fine structure splitting, stress, entangled photons}

\begin{abstract}

Eliminating the fine structure splitting (FSS) of excitons in self-assembled
quantum dots (QDs) is essential to the generation of high quality
entangled photon pairs. It has been shown that the FSS has a lower bound 
under uniaxial stress. In this letter, we 
show that the FSS of excitons in a general
self-assembled InGaAs/GaAs QD can be fully suppressed via combined stresses
along the [110] and [010] directions. The result is confirmed by
atomic empirical pseudopotential calculations. For all the QDs we studied, the FSS
can be tuned to be vanishingly small ($<$ 0.1 $\mu$eV), which is sufficient small
for high quality entangled photon emission.
\end{abstract}

\maketitle

There has been continuous interest in finding an efficient entangled photon
source for quantum information applications. Benson {\it et al.} proposed that a
biexciton cascade process in a self-assembled quantum dot (QD) can be used to
generate the ``event-ready'' entangled photon pairs.\cite{benson00} In this
scheme, a biexciton decays into two photons via two paths of different
polarizations $|H\rangle$ and $|V \rangle$. If the two paths are
indistinguishable, the two emitted photons are polarization
entangled.\cite{stevenson06,benson00} Unfortunately, the $|H\rangle$- and
$|V\rangle$-polarized photons may have a small energy difference, known as the
fine structure splitting (FSS),\cite{gammon96,bayer02} due to the asymmetric electron-hole exchange
interaction in the QDs.\cite{bester03,he08} The FSS is typically about -40
$\sim$ +80 $\mu$eV in the InAs/GaAs QDs,\cite{young05} which is much larger than
the radiative linewidth ($\sim$ 1.0 $\mu$eV), and therefore provides ``which
way'' information about the photon decay path that destroys the photon
entanglement.\cite{stevenson06,hafenbrak07} Although, it is possible to ``cherry-pick'' a QD
that has tiny FSS from a large amount of QDs,\cite{stevenson06} it is highly
desired that the FSS can be tuned by the external fields in a controlled way.
Indeed, it has been successfully demonstrated that the FSS can be tuned by
electric fields,\cite{kowalik05,gerardot07,vogel07,marcet10,bennett10} magnetic fields,\cite{stevenson06b}
and uniaxial stress.\cite{seidl06,plumhof11}

Applying a stress is an effective way to reduce the FSS in a QD
experimentally.\cite{seidl06,plumhof11} For an ideal QD with $C_{2v}$ symmetry,
the FSS can be reduced to exact zero when the stress is applied along the [110]
direction.\cite{singh10} However, for a general QD with $C_1$ symmetry, there is
a lower bound for the FSS\cite{singh10} because the two bright states belong to
the same symmetry representation. Gong {\it et al.}\cite{gong11} derived a
general relation between the FSS, the exciton polarization angle and the
uniaxial stress. They have shown that the FSS lower bound can be predicted by
the polarization angle and the FSS under zero stress. Similar relationship is
also found when the FSS is tuned by vertical electric field,\cite{bennett10}
where the minimal FSS under electric field can be as small as 0.7 $\mu$eV.
However, for most of the dots, the minimal FSS under vertical electric field is
still larger than the required splitting to observe entangled photons emission
(see Supplementary Information in Ref.~\onlinecite{bennett10}, where 16 of the 22
studied dots have the minimum FSS above 2.5 $\mu$eV). One may ask whether we can
further reduce the FSS of a {\it general} QD by the external electric fields or
stresses, to nearly zero.

In this Letter, we demonstrate that indeed the FSS of a general self-assembled
InGaAs/GaAs QDs with $C_1$ symmetry can be tuned to be vanishingly small by
applying combined stresses along the [110] and [010] directions. The results are
confirmed by atomistic empirical pseudopotential
calculations.\cite{williamson00,bester-review} The minimal FSS is generally below 0.1 $\mu$eV,
which is sufficient for high quality on-demand entangled photon emissions.

\begin{figure}
	\centering
	\includegraphics[width=3.0in]{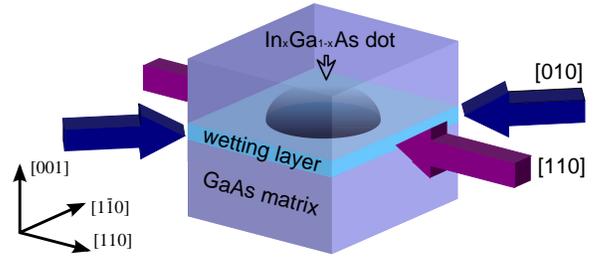}
	\caption{The InGaAs QD is embedded in a $60 \times 60 \times 60$ GaAs
	matrix. External stresses are applied along the $[010]$ and $[110]$
	directions simultaneously.}
	\label{fig:QD-under-stress}
\end{figure}

In a previous work, Gong {\it et al.} proposed an exciton Hamiltonian under uniaxial stress,\cite{gong11}
\begin{equation}
H({\bf n}, p) = H_{2v} + V_{1} + V_s({\bf n}) p \, ,
\end{equation}
where ${\bf n}$ is the external stress direction, and $p$ is the magnitude of the stress.
$H_{2v}$ represent the Hamiltonian of an ideal QD with $C_{2v}$ symmetry,
whereas $V_1$ lower the dot symmetry to $C_1$,
due to local structure deformations, alloy distribution\cite{mlinar09} and interfacial
effects\cite{bester05a} etc. $V_s({\bf n})p$ is the potential change due to
the external stress. Note that the lattice deformation under external stress($< 10^2$ MPa) is
generally smaller then 0.1\%, thus in the above Hamiltonian the term $O(p^2)$ is
omitted. The eigenvectors of the two bright states of $H_{2v}$ are
$|3\rangle = |\Gamma_2\rangle + i|\Gamma_4\rangle$ and
$|4\rangle = |\Gamma_2\rangle - i|\Gamma_4\rangle$, with corresponding eigenvalues $E_3$
and $E_4$, respectively. In the absence of an in-plane magnetic field, the coupling between the
dark states and bright states is negligible. We therefore write the Hamiltonian in the space spanned
by the two bright states of $H_{2v}$,
\begin{equation}
H = \begin{pmatrix} \bar{E} + \delta +\alpha_3 p & \kappa +\beta p \\
						  \kappa +\beta p & \bar{E} - \delta + \alpha_4 p
\end{pmatrix},
\label{eq-22}
\end{equation}
where $\bar{E}+ \delta = \bra{3}H_{2v}+V_1\ket{3}$, $\bar{E} - \delta =\bra{4}H_{2v}+V_1\ket{4}$.
$\alpha_i = \bra{i}V_s({\bf n})\ket{i}$ ($i$=3, 4), $\kappa = \bra{3}V_{1}\ket{4}$ and
$\beta = \bra{3}V_s({\bf n})\ket{4}$. Because the Hamiltonian has a time-reversal symmetry,
all parameters can therefore be set to real values for simplicity.
We define $\alpha = \alpha_3 - \alpha_4$ and $\gamma = \alpha_3
+ \alpha_4$. The eigenstates of the two bright states are the linear combination of
states, $\ket{3}$, $\ket{4}$ i.e., $\ket{\psi_{-}}= \cos\theta \ket{3} + \sin\theta \ket{4}$ and
$\ket{\psi_{+}}= -\sin\theta \ket{3} + \cos\theta \ket{4}$.
The FSS $\Delta(p)$ for QDs under uniaxial stress $p$ is,
\begin{equation}
\Delta (p) = \sqrt{4(\beta p+\kappa)^2 + (\alpha p + 2\delta)^2}\, .
\label{eq-FSS}
\end{equation}
According to the symmetry analysis (and verified by the atomistic theory), 
one has $\beta$ = 0 when the stress is applied along the [110] ([1$\bar{1}$0])
direction, and $\alpha$=0, if the stress is applied
along the [010] ([100]) direction.\cite{gong11}
Since the uniaxial stress can not make ($\beta p + \kappa$) and $(\alpha p + 2\delta)$
equal to zero simultaneously, there is always a lower bound for FSS when
$\kappa\delta \ne 0$,
which is determined by 2$|\kappa|$ or 2$|\delta|$ depending on the direction
 of the stress.

The above results suggest that diagonal and off-diagonal terms of the Hamiltonian
matrix Eq.~(\ref{eq-22}) can be tuned by the stress along the [110] and [010]
directions independently. Therefore, it is possible to further reduce the FSS
by applying the stresses along the [110] and [010] directions simultaneously.
To show this, we generalize the 2$\times$2 model exciton Hamiltonian under
uniaxial stress to that under the combined stresses,
\begin{equation}
H = \begin{pmatrix} \bar{E} + \delta +\alpha_3 p_{[110]} & \kappa +\beta p_{[010]} \\
\kappa +\beta p_{[010]} & \bar{E} - \delta + \alpha_4 p_{[110]}
\end{pmatrix} \, .
\label{eq-22-new}
\end{equation}
The FSS for QDs under the combined stress is
\begin{equation}
	\Delta(p)= \sqrt{4(\beta p_{[010]} + \kappa)^2
		+ (\alpha p_{[110]} + 2\delta)^2} \, .
\label{eq-fss-new}
\end{equation}
The polarization angle $\theta$ vs $p_{[110]}$ and $p_{[010]}$
can be calculated from,
\begin{equation}
\tan(\theta_{\pm}) = %c_3/c_4 \, .
\frac{ -2\delta -\alpha p_{[110]} \mp \Delta(p)}{2(\beta p_{[010]}+\kappa)}
\label{eq-theta}
\end{equation}
Naively, one may expect that by choosing $p_{[010]}=-\kappa /\beta$ and
$p_{[010]}=-2\delta /\alpha$, the FSS can be tuned to be exactly zero.  However, it
should be note that the exact zero FSS is forbidden by the symmetry restriction
for a QD with $C_1$ symmetry. In the above model, we neglect the higher order
terms, which may become important when FSS becomes very small. To see what is
the minimal FSS one can obtain under the combines stresses, we perform atomistic
empirical pseudopotential calculations\cite{williamson00,bester-review} on the FSS under the
combined stressed along the [010] and [110] directions.  Remarkably, we find that
the FSS of a general InGaAs/GaAs QD can be reduced to be vanishingly small ($<$0.1
$\mu$eV) under the combined stresses.

The InGaAs/GaAs QDs are modeled by embedding the InGaAs dots in a 60$\times$
60$\times$ 60 8-atom GaAs supercell.  The QDs are assumed to be grown along the
[001] direction, on the top of the one monolayer InAs wetting layers.  We apply
combined stresses along the [110] and the [010] directions as shown in
Fig.~\ref{fig:QD-under-stress}. The stresses can be either compressive or
tensile depending on the structure of QDs. The resultant deformation of the
supercell under stresses is approximated by the linear superposition of the
deformation of the cell under individual uniaxial stresses. Optimal atomic
positions under stresses are obatained with valance force field
method.\cite{keating66,martin70} We solve the Schr\"{o}dinger equation to obtain
the single-particle energy levels and wavefunctions using a strained linear
combination of Bloch band method.\cite{wang99b} The exciton energies are then
calculated via the many-particle configuration interaction
method,\cite{Franceschetti99} in which the exciton wave functions are expanded
in Slater determinants constructed from all confined electron and hole
single-particle states.

\begin{figure}%[htb]
	\centering
	\includegraphics[width=3.2in]{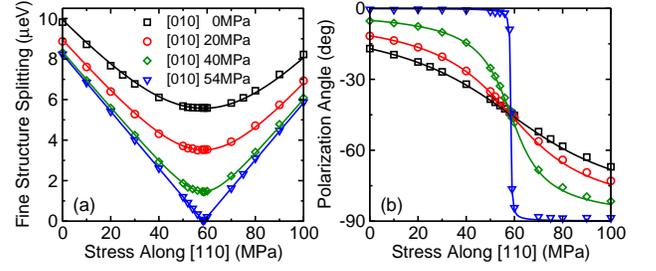}
	\caption{Change of FSS (a) and polarization angle (b)
	under the combined stress along the $[110]$ and $[010]$ directions
	for the lens-shaped In$_{0.6}$Ga$_{0.4}$As/GaAs dot.
	The scatters are the results of numerical calculations 
    whereas the solid lines are the results of the two-level model.}
	\label{fig:fss-theta}
\end{figure}

\begin{table*}
\caption{\label{tab:param}Parameters for In$_x$Ga$_{1-x}$As/GaAs QDs with
different shape and compositions under combined stresses along the [110] and
[010] directions. The unit of dot diameter $D$ and hight $h$ is nm. The fine
structure $\Delta(p)$ is in unit of $\mu$eV and the combined critical stresses
$p_c=(p_{[110]},p_{[010]})$ are in unit of MPa.  }
\begin{tabular}{m{12em}ccccccc}
\hline\hline
    & \multicolumn{2}{c}{Pesudopotential Calculation} & \multicolumn{5}{c}{Model Parameters and Predictions}\\
\cline{2-3}\cline{4-8}
QDs & $\Delta(p_c)$ & $p_c$ & $\alpha$ & $\delta$ & $\beta$ & $\kappa$ & $p_c$ \\ \hline
Cone ($x=0.6$)\newline $D=25$, $h=3.5$ &
 0.084 &(58.0, -36.0) & -0.15 & 4.36 & 0.06 & 2.04 & (58.6, -36.1)\\  \hline
Truncated Cone ($x=0.6$)\newline $D_{B}=25$ \newline $D_{T}=20$, $h=3.5$ &
 0.032 &(11.3, -45.7) & -0.14 & 0.79 & 0.05 & 2.23 & (11.5, -46.0) \\  \hline
Elongated Lens ($x=0.6$)\newline $D_{[110]}=20$ \newline $D_{[1\bar{1}0]}=25$, $h=3.5$ &
 0.045 &(75.8, -19.8) & -0.17 & 6.37 & 0.06 & 1.24 & (76.0, -20.1) \\  \hline
Elongated Lens ($x=0.6$)\newline $D_{[110]}=25$ \newline $D_{[1\bar{1}0]}=20$, $h=3.5$ &
 0.076 &(-36.8, -10.7) & -0.17 & 3.03 & 0.06 & 0.70 & (-36.6, -11.4) \\  \hline
Lens ($x=0.8$)\newline $D=25$ $h=3.5$ &
 0.048 &(35.6, 14.0) & -0.19 & 3.46 & 0.06 & -0.89 & (35.8, 14.1) \\  \hline
Lens ($x=0.6$)\newline $D=25$ $h=3.5$ &
 0.039 &(58.5, 54.0) & -0.14 & 4.11 & 0.05 & -2.79 & (58.5, 54.3) \\  \hline
Lens ($x=0.6$)\newline $D=25$ $h=5.0$ &
 0.083 &(-12.4, -43.6) &-0.13 & -0.79 & 0.045 & 1.99 & (-12.2, -43.8)\\   \hline
Pyramid ($x=0.6$)\newline $D=25$ $h=3.5$ &
 0.075 &(27.5, 1.6) &-0.13 & 1.77 & 0.051 & -0.092 & (28.0, 1.78)\\
\hline\hline
\end{tabular}
\end{table*}

We have calculated 8 InAs/GaAs QDs with different geometries, including
(elongated-)lens, pyramid, truncated cone, etc. and different alloy
compositions. Here we take a lens-shaped In$_{x}$Ga$_{1-x}$As/GaAs dot with $x$
= 0.6, diameter $D$ = 25 nm, and height $h$= 3.5 as an example. We obtain
similar results for all other QDs. Fig.~\ref{fig:fss-theta} depicts the
behavior of the FSS and polarization angle of the dot under combined [110] and
[010] stresses. We plot the FSS and exciton polarization angles as functions of
stress along the [110] direction under $p_{[010]}$=0, 20, 40 and 54 MPa. For a
given stress along the [010] direction, the FSS goes through an anti-crossing as a
function of stress along the [110] direction, which is very similar to the
results under uniaxial stress.\cite{gong11} The critical stress along the [110]
direction changes little with the stress applied along the [010] direction.
However the minimal FSS changes dramatically for different stresses applied
along the [010] direction, as we expected. When no stress is applied, the FSS of
the QD is approximately 9.8 $\mu$eV. When we apply the stress along the [110]
direction, one can reduce the FSS to approximately 5.6 $\mu$eV, and 8.2 $\mu$eV
if the stress is applied along the [010] direction. In either case, the FSS is
much larger than 1 $\mu$eV to suitable for high quality entangled photon
emitters. However, when combined stresses along the [110] and [010] directions
are applied, the lowest FSS we obtain is 0.04 $\mu$eV at $p_{[110]}$=58.5 MPa
and $p_{[010]}$=54 MPa. The polarization angles rotate clockwise or
anticlockwise depending on the magnitude of stress along the [010] direction and
the polarization angle at the critical stress always tends to be $-45^{\circ}$ or
$45^{\circ}$. The calculated FSS and exciton polarization angles are in
excellent agreement with model predictions. The minimal FSS of all other QDs
are also $<$0.1 $\mu$eV under the combined stresses (See Table~\ref{tab:param}
for critical stresses and the FSS minimum of all dots we studied).
The tunable area in which FSS is less than 1 $\mu$eV is proportional to
$2/\alpha$ and $1/\beta$. In typical InGaAs/GaAs dots, $|\alpha|\sim$ 0.1-0.4
$\mu$eV/MPa and $|\beta|\sim$ 0.04-0.1 $\mu$eV/MPa,\cite{gong11} therefore the
tunable area for FSS less than 1 $\mu eV$ is approximately $5\times 10$~MPa$^2$,
which is easy to control in experiments.
 
When the stresses along the two directions approach the critical point at the
same time, the FSS becomes very small, the higher order terms in
Eq.~(\ref{eq-22-new}) may not be ignored, and therefore the FSS can never be
exactly zero.  Such higher order terms are fully included in the atomistic
calculation. The comparison between the model predictions and exact numerical
simulations can reveal the role of these terms. As shown in Fig.~\ref{fig:fss-theta},
the numerical results (scatters) and two-level model predictions
(solid lines), using the parameters fitted from the results of uniaxial
stress, are in excellent agreement. This suggests that the higher order terms
are indeed unimportant except extremely close to the critical stresses.

\begin{figure}[htb]
	\centering
	\includegraphics[width=3.2in]{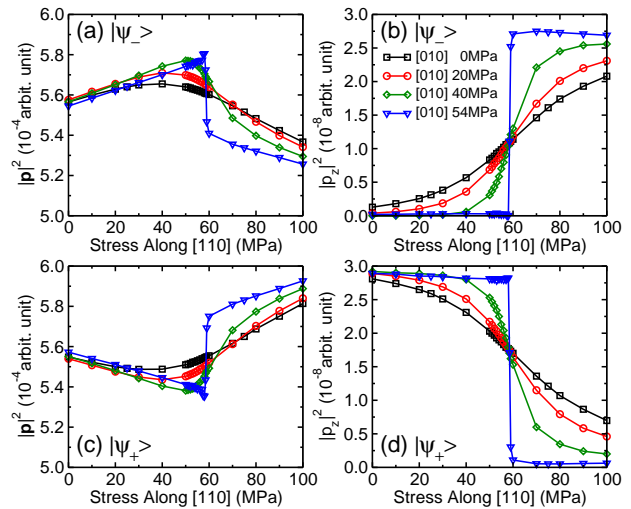}
	\caption{Left panel: The change of the total dipole moment under the
	stresses for the two bright states: $|\psi_-\rangle$ (a) and $|\psi_+\rangle$ (c), in
	the lens-shaped In$_{0.6}$Ga$_{0.4}$As/GaAs dot.  Right panel: The change of
	the $z$-component of dipole moment under combined stresses for the two
	bright states: $|\psi_-\rangle$ (b) and $|\psi_+\rangle$ (d).}
	\label{fig:dipole}
\end{figure}

Figure \ref{fig:dipole}(a), \ref{fig:dipole}(c) depict the total dipole moment 
$|{\bf p}|^2$=$p_x^2$+$p_y^2$+$p_z^2$ of the
two bright states under combined stresses. The total dipole moments of the two
states change a little in opposite directions under the stresses, and the
lifetimes of the excitons change accordingly. 
However, the sum of the total dipole moments of the two bright
states shows very little change ($<$ 1\%) for stresses up to 100 MPa.
Interestingly, for applied stress $p_{[010]}=54$ MPa (critical stress
  along the [010] direction), the total dipole moments of the two states 
change suddenly near the critical stress $p_{[110]}=58$ Mpa. This is because, at 
$p_{[010]}=54$ MPa, the dipole moments of the two states 
are approximately aligned along the $[110]$ and $[1\bar{1}0]$ directions,
with one increases whereas one decreases under the stress 
along the [110] direction. When the stress approach the critical stress along
the [110] direction, the polarization angle rotates very fast 
from $0^{\circ}$ to $-90^{\circ}$, i.e., the two states switch polarization,
which causes the jump of the total dipole moments of the two states.
There are also very small $z$ components of the dipole moment $p_z$
because of the valence band mixing,
as shown in Fig. \ref{fig:dipole}(b), \ref{fig:dipole}(d). 
Therefore,  when we project the dipole moment into the x-y plane, 
the polarization of the two bright
states are not exactly orthogonal to each other. The deviation is
approximately 1$^\circ$ - 2$^\circ$.

Bennett {\it et al.} have successfully tuned the
FSS by applying a vertical electric field.\cite{bennett10} 
According to the symmetry analysis,
the effect of a vertical electric field is equivalent to the uniaxial stress
along the [110] direction. Therefore, it is also possible to tune the FSS to
approximately zero by applying a combined vertical electric field and an
uniaxial stress along the [010] direction, which might be more feasible
experimentally. We leave this for future study.

To summarize, we have demonstrated that the FSS in general self-assembled QDs
can be reduced to nearly zero under combined stresses along the $[110]$ and
$[010]$ directions. A wide range of stresses can reduce the FSS to less than 1
$\mu$eV and thus provide great merit in experiments.

LH acknowledges the support from the Chinese National
Fundamental Research Program 2011CB921200,
National Natural Science Funds for Distinguished Young Scholars,
and the Fundamental Research Funds for the Central Universities
No. WK2470000006.

%Reference

%\bibliography{ref}

%merlin.mbs aipnum4-1.bst 2010-07-25 4.21a (PWD, AO, DPC) hacked
%Control: key (0)
%Control: author (8) initials jnrlst
%Control: editor formatted (1) identically to author
%Control: production of article title (-1) disabled
%Control: page (0) single
%Control: year (1) truncated
%Control: production of eprint (0) enabled
%

\end{document}